\documentclass[10pt,twocolumn]{article}
\usepackage{wrapfig}
\usepackage{graphicx}
\usepackage{amsmath,amssymb}

\title{\textbf{Theoretical background for observing ultra-slow microwaves
     in a Bose-Einstein condensate of alkali atoms}}

\author{Yuriy V. Slyusarenko and Andrey G. Sotnikov
    \vspace{3mm}
    \\
\emph{\small Akhiezer Institute for Theoretical Physics, NSC KIPT,
                1~Akademicheskaya str., 61108 Kharkiv, Ukraine}}
\date{}

\begin{document}
\twocolumn[\maketitle \small{
           We represent a new microscopic approach that allows
           studying the propagation properties of microwaves in a
           Bose-Einstein condensate of alkali atoms. It is assumed that
           the frequency of signal is tuned up to the transition
           between hyperfine ground state levels of such atoms.
           Pulse slowing conditions dependence on the system parameters is found.
           It is shown that the slowed signal can propagate in
           mentioned system with rather small energy loss.
           Such phenomenon is also studied in case of hyperfine levels
           Zeeman splitting. A possibility of ultra-slow microwaves observing
           in a condensed gas of cesium atoms is discussed.
           \vspace{3mm}
    \vspace{4mm}
    \\}]

\section{Introduction}
In 1999 the unique opportunity to observe the ultra-slow pulses of
the optical region of spectra in a Bose-Einstein condensate (BEC) of
sodium atoms was shown \cite{ref.1}. In this and subsequent
experiments laser pulses were tuned up to dipole-allowed
transitions.

Naturally, a question arises about the possibility of using another
levels for observing electromagnetic pulses slowing in BEC of alkali
atoms. In our opinion, it seems to be convenient to use microwave
pulses tuned up to the transition between ground state hyperfine
structure levels of hydrogen-like atoms. It is easy to see that such
levels are rather stable. This fact, in particular, explains the
universe abundance of microwaves with "famous" 21~cm wavelength,
which correspond to the transitions between hyperfine states of the
atomic hydrogen. Moreover, there are many trapping-related
experiments with ultra-cold atoms, which are prepared in different
hyperfine states (see, \textit{e.g.}, experiments for the
multi-component BEC observing~\cite{ref.2}).

However, as it is easy to see, alkali atoms in the ground state do
not have a dipole moment. As a result, theoretical descriptions
(see, \textit{e.g.}, ref.~\cite{ref.3}), in which atoms are
considered as dipoles, become no longer convenient. For our opinion,
the most appropriate way to go out from this situation is using a
microscopic approach. Such an approach, for example, was developed
in ref.~\cite{ref.4}. It is based on the approximate formulation of
the second quantization method and takes into account the presence
of bound states of particles. As it is shown below, such method is
universal in some respect and allows to describe the response of the
system not only to microwave range perturbation (the signal tuned up
to the transitions between hyperfine ground state levels), but also
to the optical range perturbation (the pulse tuned up to the
dipole-allowed transitions)~\cite{ref.5}.

Let us briefly state out some basic principles of the suggested
approach.

\section{Theoretical basis}
As one can suggest, alkali atoms are similar to the hydrogen atom by
the internal structure, and can be considered as bound states of two
"elementary" particles (valence electron and atomic core). From the
standpoint of the quantum mechanics, it is also known that a bound
state energy can lie not only in the discrete spectrum region, but
also in the continuum one. This fact was taken into account in
ref.~\cite{ref.4} in construction of novel formulation of the second
quantization method in the presence of bound states (atoms). It was
also used in ref.~\cite{ref.5} in construction of microscopic theory
of the hydrogen-like low temperature plasma response to the external
electromagnetic field.

Thus, to construct a microscopic approach one needs to consider a
system, that consists of two kinds of free fermions (ions and
electrons) and their bound states (alkali atoms), as it was done in
ref.~\cite{ref.4}. But, to simplify the further description, we have
to note the following. In recent theoretical investigations
\cite{ref.6} of hydrogen-like low temperature plasma in an
equilibrium state it was shown that the density of free fermions is
exponentially small in comparison with the density of atoms.

Therefore, to build microscopic approach describing electrodynamic
processes at low temperatures one can use the mentioned second
quantization method with preserving the quantities that correspond
only to bound states (atoms) contribution. According to
ref.~\cite{ref.4}, the system Hamiltonian can be written in the
following form:
\begin{equation}\label{eq.1}                                    %%%% ---- %%%% eq 1
    \hat{\mathcal H}(t)
    =\hat{\mathcal H}_0
    +\hat{\mathcal H}_{\text{int}}
    +\hat{V}(t),\quad
    \hat{\mathcal H}_0
    =\hat{\mathcal H}_{\text{ph}}
    +\hat{\mathcal H}_{\text{p}},
\end{equation}
where $\hat{\mathcal H}_{\text{ph}}$ is the Hamiltonian for free
photons and $\hat{\mathcal H}_{\text{int}}$ is the Hamiltonian of
interaction between atoms (we neglect Hamiltonians~$\hat{\mathcal
H}_{\text{ph}}$ and $\hat{\mathcal H}_{\text{int}}$ below, but for
the explicit form see also ref.~\cite{ref.4}).

The operator $\hat{\mathcal H}_{\text{p}}$ in eq.~(\ref{eq.1}) is
the Hamiltonian for free particles (atoms)
\begin{equation}\label{eq.2}                                    %%%% ---- %%%% eq 2
\begin{split}
    \hat{\mathcal H}_{\text{p}}
    =\sum_{\alpha}\int d\textbf{X}
    \biggl\{
    \dfrac{1}{2m}
    \dfrac{\partial{\hat\eta}_{\alpha}^{\dag}(\textbf{X})}
    {\partial\textbf{X}}
    \dfrac{\partial{\hat\eta}_{\alpha}
    (\textbf{X})}{\partial\textbf{X}}\biggr.
    \\
    \biggl.+\varepsilon_{\alpha}{\hat\eta}_{\alpha}^{\dag}
    (\textbf{X}){\hat\eta}_{\alpha}(\textbf{X})
    \biggr\},
\end{split}
\end{equation}
where ${\hat\eta}_{\alpha}^{\dag}(\textbf{X})$, ${\hat\eta}_{\alpha}
(\textbf{X})$ are the creation and annihilation operators of
hydrogen-like (alkali) atoms with the set of quantum numbers
$\alpha$ at the point $\textbf{X}$; $\varepsilon_{\alpha}$ is the
energy of an atom in the state characterized by quantum numbers
$\alpha$; $m$ is the atomic mass.

And, finally, the operator $\hat{V}(t)$ in eq.~(\ref{eq.1})
represents the Hamiltonian that describes the interaction of
particles with the electromagnetic field
\begin{equation*}
    \hat{V}(t)
    =-\dfrac{1}{c}\int d\textbf{x}\textbf{A}^{(e)}(\textbf{x},t)
    \hat{\textbf{j}}(\textbf{x})
    +\int d\textbf{x}\varphi^{(e)}(\textbf{x},t)
    \hat{\sigma}(\textbf{x}),
\end{equation*}
where $\textbf{A}^{(e)}(\textbf{x},t)$ and
$\varphi^{(e)}(\textbf{x},t)$ are the vector and scalar potentials
of the external electromagnetic field, respectively, and operators
$\hat{\textbf{j}}(\textbf{x})$ and $\hat{\sigma}(\textbf{x})$ are
the charge and current density operators, respectively:
\begin{equation}\label{eq.3}                               %%%% ---- %%%% eq 3
\begin{split}
    &\hat{\sigma}(\textbf{x})=\dfrac{1}{\mathcal{V}}
    \sum\limits_{\textbf{p},\textbf{p}'}
    \sum\limits_{\alpha,\beta}
    e^{i\textbf{x}(\textbf{p}'-\textbf{p})}
    \\
    &\qquad\qquad\times
    \sigma_{\alpha\beta}(\textbf{p}-\textbf{p}')
    \hat{\eta}_{\alpha}^{\dag}(\textbf{p})
    \hat{\eta}_{\beta}(\textbf{p}'),
    \\
    &\hat{\textbf{j}}(\textbf{x})=\dfrac{1}{\mathcal{V}}
    \sum\limits_{\textbf{p},\textbf{p}'}
    \sum\limits_{\alpha,\beta}
    e^{i\textbf{x}(\textbf{p}'-\textbf{p})}
    \biggl(\textbf{I}_{\alpha\beta}(\textbf{p}-\textbf{p}')
    \biggr.
    \\
    &\qquad\biggl.
    +\dfrac{(\textbf{p}+\textbf{p}')}{2M}
    \sigma_{\alpha\beta}(\textbf{p}-\textbf{p}')
    \biggr)
    \hat{\eta}_{\alpha}^{\dag}(\textbf{p})
    \hat{\eta}_{\beta}(\textbf{p}').
\end{split}
\end{equation}
Here $\mathcal{V}$ is the system volume. Note that the charge and
current density matrix elements in eq.~(\ref{eq.3}) can be expressed
(see ref.~\cite{ref.4}) in terms of atom wave functions
$\varphi_{\alpha}(\textbf{x})$:
\begin{equation}\label{eq.4}                              %%%% ---- %%%% Eq. 4
\begin{split}
    &\sigma_{\alpha\beta}(\textbf{k})
    =e\int d\textbf{y}\varphi_{\alpha}^{*}(\textbf{y})
    \varphi_{\beta}(\textbf{y})
    \\
    &\qquad\times\left[\exp{\left(i\dfrac{m_{\text{p}}}{m}
    \textbf{k}\textbf{y}\right)}
    -\exp{\left(-i\dfrac{m_{\text{e}}}{m}
    \textbf{k}\textbf{y}\right)}\right],
    \\
    &\textbf{I}_{\alpha\beta}(\textbf{k})
    =-\dfrac{ie}{2}\int d\textbf{y}
    \\
    &\qquad\times\left(\varphi_{\alpha}^{*}(\textbf{y})
    \dfrac{\partial\varphi_{\beta}(\textbf{y})}
    {\partial\textbf{y}}-
    \dfrac{\partial\varphi_{\alpha}^{*}(\textbf{y})}
    {\partial\textbf{y}}\varphi_{\beta}(\textbf{y})
    \right)
    \\
    &\quad\times
    \left[\dfrac{1}{m_{\text{p}}}\exp{\left(i\dfrac{m_{\text{e}}}{m}
    \textbf{k}\textbf{y}\right)}
    +\dfrac{1}{m_{\text{e}}}\exp{\left(-i\dfrac{m_{\text{p}}}{m}
    \textbf{k}\textbf{y}\right)}\right],
\end{split}
\end{equation}
where $e$ is the electron charge absolute value, $m_\text{p}$ and
$m_\text{e}$ are the atomic core and electron mass, respectively
($m=m_\text{p}+m_\text{e}$).

\section{System response to the external electromagnetic field in
 the framework of Green functions formalism}
Non-relativistic equations of quantum electrodynamics, which were
found in ref.~\cite{ref.4} on the basis of
Hamiltonians~(\ref{eq.1})-(\ref{eq.4}) and which, in turn, were used
in ref.~\cite{ref.5}, allow to study the linear response of the
system to a perturbation by the external electromagnetic
field~\cite{ref.7}. To find out such a response it is most
convenient to use the Green functions formalism (as it was done in
ref.~\cite{ref.5}). In the framework of this formalism the scalar
(charge) Green function can be defined as:
\begin{equation*}
    G^{(+)}(\textbf{x},t)=-i\theta(t)\text{Sp}
    {w[\hat{\sigma}(\textbf{x},t),\hat{\sigma}(0)]},
\end{equation*}
where $\theta(t)$ is the Heaviside function, $w$ is the Hibbs
distribution operator and charge density operators (see
definition~(\ref{eq.3})) must be taken in the Heisenberg
representation. Neglecting the interaction between particles (as it
can be done for dilute gases), one can get the expression for the
scalar Green function Fourier transform (see ref.~\cite{ref.5}):
\begin{equation}\label{eq.5}                              %%%% ---- %%%% Eq. 5
\begin{split}
    G^{(+)}(\textbf{k},\omega)
    &=\dfrac{1}{\mathcal{V}}
    \sum\limits_{\textbf{p}}
    \sum\limits_{\alpha,\beta}
    \sigma_{\alpha\beta}(\textbf{k})
    \sigma_{\beta\alpha}(-\textbf{k})
    \\
    &\qquad\times
    \dfrac{f_{\alpha}(\textbf{p}-\textbf{k})
    -f_{\beta}(\textbf{p})}
    {\varepsilon_{\alpha}(\textbf{p})-
    \varepsilon_{\beta}(\textbf{p}-\textbf{k})
    +\omega+i0},
\end{split}
\end{equation}
where
$\varepsilon_{\alpha}(\textbf{p})=\varepsilon_{\alpha}+\textbf{p}^{2}/{2m}$,
$f_{\alpha}(\textbf{p})$ is the bosonic distribution function of the
ideal gas of hydrogen-like (alkali) atoms
\begin{equation*}
    f_{\alpha}(\textbf{p})=
    \{\exp[(\varepsilon_{\alpha}(\textbf{p})
    -\mu_{\alpha})/T]-1\}^{-1}.
\end{equation*}
Here $\mu_{\alpha}$ is the chemical potential of atoms in the state
$\alpha$, $T$ is the temperature of the gas that is taken in energy
units.

Using the developed theory, it is not difficult to find the
permittivity of such gas at low temperatures. From eq.~(\ref{eq.5})
one gets (see also ref.~\cite{ref.5}):
\begin{equation}\label{eq.6}                              %%%% ---- %%%% Eq. 6
\begin{split}
    \epsilon^{-1}(\textbf{k},\omega)
    =1+\dfrac{4\pi}{k^2}
    \dfrac{1}{\mathcal{V}}
    \sum\limits_{\textbf{p}}
    \sum\limits_{\alpha,\beta}
    \sigma_{\alpha\beta}(\textbf{k})
    \sigma_{\beta\alpha}(-\textbf{k})
    \\
    \times
    \dfrac{f_{\alpha}(\textbf{p}-\textbf{k})
    -f_{\beta}(\textbf{p})}
    {\varepsilon_{\alpha}(\textbf{p})-
    \varepsilon_{\beta}(\textbf{p}-\textbf{k})
    +\omega+i0}.
\end{split}
\end{equation}

As it is known, at extremely low temperatures a Bose-Einstein
condensate (BEC) of alkali atoms can be formed. At temperatures much
lower the critical point temperature $T\ll T_0$ the atomic
distribution functions $f_\alpha (\textbf{p})$ are proportional to
the Dirac delta-function $\delta(\textbf{p})$. Note also that at
temperatures $T\lesssim T_0$ the chemical potential~$\mu_{\alpha}$
does not depend on temperature and must be set equal to energy of
the lowest level of atoms, which form BEC (see
refs.~\cite{ref.6,ref.8}). Therefore, after integration of
eq.~(\ref{eq.6}) over momentum $\textbf{p}$ the expression for the
permittivity of the studied gas in BEC state ($T\rightarrow0$) takes
the form:
\begin{equation}\label{eq.7}                              %%%% ---- %%%% Eq. 7
\begin{split}
    \epsilon^{-1}(\textbf{k},\omega)
    \approx 1&+\dfrac{1}{2\pi^2 k^2}
    \sum\limits_{\alpha,\beta}
    \sigma_{\alpha\beta}(\textbf{k})
    \sigma_{\beta\alpha}(-\textbf{k})
    \\
    &\times\left[
    \dfrac{\nu_{\alpha}}
    {\omega+\Delta\varepsilon_{\alpha\beta}
    -\varepsilon_{\text{k}}+i\gamma_{\alpha\beta}}\right.
    \\
    &\qquad\left.-\dfrac{\nu_{\beta}}
    {\omega+\Delta\varepsilon_{\alpha\beta}
    +\varepsilon_{\text{k}}+i\gamma_{\alpha\beta}}\right],
\end{split}
\end{equation}
where $\nu_\alpha$ is the density of condensed atoms in the quantum
state $\alpha$, $\varepsilon_{\text{k}}=k^2/2M$, and the quantities
$\sigma_{\alpha\beta}(\textbf{k})$ are still defined by the
formula~(\ref{eq.4}). Note that due to damping processes in real
systems we also introduce the linewidth~$\gamma_{\alpha\beta}$,
concerned with the transition probability between the state $\alpha$
and state $\beta$. As it easy to see, in eq.~(\ref{eq.7}) at
frequencies close to the energy interval
$\Delta\varepsilon_{\alpha\beta}$
($\Delta\varepsilon_{\alpha\beta}\equiv\varepsilon_{\alpha}
-\varepsilon_{\beta}$) some peculiarities appear. In fact, such a
behavior must have a strong impact on the dispersion characteristics
of the gas and can be an underlying condition for pulses slowing.

To study the propagation properties of the signal one should add the
dispersion relation for free waves, that can spread in the studied
system:
\begin{equation}\label{eq.8}                              %%%% ---- %%%% Eq. 8
    {\omega^{2}\over
    c^{2}}\epsilon(\textbf{k},\omega)
    \mu(\textbf{k},\omega)-k^{2}=0,
\end{equation}
where $\mu(\textbf{k},\omega)$ is the magnetic permeability that can
be found analogically in the framework of Green functions formalism
(see in that case ref.~\cite{ref.5}).

\section{Microwaves slowing caused by hyperfine structure levels}
In the previous section the expression for the permittivity of the
system in BEC state (see eq.~(\ref{eq.7})) was represented. For the
system with the frequency of the external field tuned up to the
difference between two defined levels (marked below by subscripts 1
and 2) it can be written in a more suitable form:
\begin{equation}\label{eq.9}                              %%%% ---- %%%% Eq. 9
    \epsilon^{-1}(\textbf{k},\omega)
    \approx1+\dfrac{g_1 g_2|\sigma_{12}
    (\textbf{k})|^2}{2\pi^2 k^2}
    \dfrac{(\nu_{1}-\nu_{2})}
    {\delta\omega+i\gamma}.
\end{equation}
Here $g_j$ is the degeneracy order over total spin of $j$ level
($j=1,2$), $\delta\omega=\omega-\Delta\varepsilon_{21}$ is the laser
detuning (${|\delta\omega|\ll\Delta\varepsilon_{21}}$),
$\gamma\equiv\gamma_{12}$ is linewidth related to the transition
probability from the upper to lower state. Note that we also
neglected the term $\varepsilon_{k}$ (see eq.~(\ref{eq.7})), the
substantiation of such operation is discussed below.

To study the propagation properties it is more convenient to turn to
the refractive index and the damping factor quantities. To do it we
set the magnetic permeability in the dispersion
relation~(\ref{eq.8}) close to unity, $\mu(\textbf{k},\omega)=1$. In
this case the refractive index $n(\textbf{k},\omega)$ and the
damping factor $\chi(\textbf{k},\omega)$ can be expressed in terms
of the real and imaginary part of the permittivity ($\epsilon'$ and
$\epsilon''$, respectively), which in turn can be derived from
eq.~(\ref{eq.9}):
\begin{equation}\label{eq.10}                              %%%% ---- %%%% Eq. 10
    \epsilon'=\dfrac{\delta\omega
    (\delta\omega+a)+\gamma^2}
    {(\delta\omega+a)^2+\gamma^2},\quad
    \epsilon''=\dfrac{\gamma a}
    {(\delta\omega+a)^2+\gamma^2},
\end{equation}
where
\begin{equation}\label{eq.11}                              %%%% ---- %%%% Eq. 11
    a(\textbf{k})=(\nu_{1}-\nu_{2})\dfrac{g_1 g_2
    |\sigma_{12}(\textbf{k})|^2}{2\pi^2 k^2}.
\end{equation}

Now one can find the dependence of group velocity on system
parameters. As it is known, the group velocity of a propagating
pulse can be defined as:
\begin{equation*}                                          %%%% ---- %%%%
    v_{\text{g}}=\dfrac{c}
    {n+\omega(\partial n/\partial \omega)}.
\end{equation*}
Thus, after some mathematical transformations, in case of energy
dissipation smallness and strong dispersion, one gets:
\begin{equation}\label{eq.12}                              %%%% ---- %%%% Eq. 12
    v_{\text{g}}\approx2c\dfrac
    {\left((\delta\omega+a)^2+\gamma^2
    \right)^2}
    {a\omega\left[(\delta\omega+a)^2-\gamma^2
    \right]}.
\end{equation}
This expression gives the opportunity to study the dependence of the
group velocity not only on detuning $\delta\omega$ and linewidth
$\gamma$, but also on characteristic properties of the system under
consideration that have an impact on the parameter $a$ value (see
eq.~(\ref{eq.11})).

Now let us obtain conditions, in which the ultra-slow microwaves
phenomenon for two-level system in BEC state can be observed. To
this end we proceed to the limit ${\delta\omega\rightarrow0}$. In
this case, according to eq.~(\ref{eq.10}), the real and imaginary
parts of permittivity tend to the limits:
\begin{equation*}                              %%%% ---- %%%%
    \lim\limits_{\delta\omega\rightarrow0}
    \epsilon'=\dfrac{\gamma^2}
    {\gamma^2+a^2},\quad
    \lim\limits_{\delta\omega\rightarrow0}
    \epsilon''=\dfrac{\gamma a}
    {\gamma^2+a^2}.
\end{equation*}
Hence, the dissipation smallness
condition~($\epsilon'\ll|\epsilon''|$) can be written as follows:
\begin{equation*}                              %%%% ---- %%%%
    \dfrac{|a|}{\gamma}\ll1.
\end{equation*}
Note that it is necessary also to add the slowing down (strong
dispersion) condition, which in this case (see eq.~(\ref{eq.12}))
can be written as:
\begin{equation*}                              %%%% ---- %%%%
    \dfrac{c}{v_\text{g}}\approx
    \dfrac{\Delta\varepsilon_{21}|a|}{2\gamma^2}\gg1,
\end{equation*}
For the defined system with a fixed energy structure
($\Delta\varepsilon,\gamma=\text{const}$) the only parameter that
can be varied is the occupation difference $(\nu_{1}-\nu_{2})$,
which is included in the parameter $a$ (see eq.~(\ref{eq.11})).
Thus, basing on these relations we get the expression that
characterizes the region where the mentioned phenomenon can be
observed:
\begin{equation}\label{eq.13}                              %%%% ---- %%%% Eq. 13
    \dfrac{\gamma}{\Delta\varepsilon_{21}}
    \ll\dfrac{|a|}{\gamma}\ll1,
\end{equation}

By considering the example of the hyperfine levels of the ground
state for cesium atoms let us demonstrate that such region can
exist. The choice of such levels, as it is mentioned above, is
stimulated by their stability and pumping capability. Note also that
for such levels the dipole transitions are forbidden, thus
transitions come from the higher order effects that result in
extremely small values of linewidths. It is shown below that such
fact gives the opportunity for a signal to propagate with a small
loss of energy.

It should be mentioned that the description can be extended to other
hydrogen-like atoms and other type of levels analogically. For
example, it can be used for a description of experiments with an
ultra-slow light in BEC of sodium atoms~\cite{ref.1}, in which the
dipole-exited states were used for a pulse slowing. In this case to
get numerical results from the developed approach one can restrict
calculations of the charge density matrix element $\sigma_{12}$~(see
eq.~\ref{eq.4}) to the first order approximation over
$(\textbf{ky})\ll1$:
\begin{equation*}                                         %%%% ---- %%%%
    \sigma^{(1)}_{12}(\textbf{k})\approx i\textbf{k}\textbf{d}_{12},
\end{equation*}
with the dipole moment $\textbf{d}_{12}$ that corresponds to the
dipole transition $1\rightarrow2$:
\begin{equation*}                                         %%%% ---- %%%%
    \textbf{d}_{12}=e\int \textbf{y}d\textbf{y}\varphi_{1}^{*}(\textbf{y})
    \varphi_{2}(\textbf{y}).
\end{equation*}
But, it is well known that alkali metals ($^{133}$Cs in particular)
in the ground state do not have a dipole moment~$\textbf{d}$, thus
to describe the pulse slowing on dipole-forbidden transitions the
charge density matrix element $\sigma_{12}$ (see
definition~(\ref{eq.4})) must be expanded to the second order over
$(\textbf{ky})\ll1$. As a result, one gets:
\begin{equation}\label{eq.14}                              %%%% ---- %%%% Eq. 14
    \sigma^{(2)}_{12}(k)\approx\dfrac{e}{3}(kr_0)^2,
\end{equation}
where $r_0$ is the atomic radius (for cesium ground state
$r_0\approx 2,6\times10^{-8}$~cm \cite{ref.9}), $e$ is the electron
charge. Taking $g_1=7$, $g_2=9$ ($g_{j}=2F_{j}+1$, $F_j$ is the
total spin of an atom in $j$ hyperfine state, $j=1,2$),
$k=(\Delta\varepsilon_{21}/c)$, where $\Delta\varepsilon_{21}\approx
3,8\times10^{-5}$~eV (microwaves with frequency 9,1926~GHz), the
linewidth $\gamma\approx3,8\times10^{-21}~\text{eV}$, corresponding
to the anticipated accuracy ($10^{-16}$) in "cesium fountain clock"
experiments \cite{ref.10}, and basing on the
expressions~(\ref{eq.11}), (\ref{eq.13}) one can find the region for
the occupation difference, in which the pulse slowing phenomenon can
be observed:
\begin{equation}\label{eq.15}                              %%%% ---- %%%% Eq. 15
    10^{-3}~\text{cm}^{-3}
    \ll|\nu_1 - \nu_2|\ll
    3\times10^{13}~\text{cm}^{-3}.
\end{equation}
From this inequality one can conclude that the effect becomes
greater with the density difference increasing until it reaches the
upper limit of the expression~(\ref{eq.15}), when damping effects
prevail in the system. We should stress that the region of
densities~(\ref{eq.15}) looks convenient from the standpoint of
experiments for BEC regime (see, \textit{e.g.}, the
experiment~\cite{ref.11}, in which cesium atoms with the density
$\nu=7\times10^{10}~\text{cm}^{-3}$ were used).

We should also note the following. As it easy to see from
eq.~(\ref{eq.12}) in the limit $\delta\omega\rightarrow0$, the sign
of the group velocity $v_{g}$ depends directly on the sign of the
quantity $a$, that in turn depends on the sign of the difference
$(\nu_1 - \nu_2)$. In other words, it depends on whether the
population is normal or inverse.

In case of normal population $(\nu_1 > \nu_2)$ the group velocity of
signal is negative. It is traditionally considered that the group
velocity for the transparent matter is positive. But here one can
conclude that due to the relation~(\ref{eq.13}), the signal can
propagate in the system with rather small dissipation
(\textit{i.e.}, in fact, the matter is transparent) and rather slow
velocity. Let us note that an observing of electromagnetic pulses
with negative group velocity is not so abnormal. The existence of
such kind of phenomena for physical systems when the wave frequency
is close to atomic (or molecular) resonances was pointed out and
studied in many works (both theoretical, see                            %% CHANGED !!! %%
\textit{e.g.}~\cite{ref.12}, and
experimental~\cite{ref.Chu82,ref.Macke85}). In case of inverse
population $(\nu_1 < \nu_2)$ more "normal" situation takes place
because the group velocity of the slowed pulse is positive.

We stress that such rather unusual phenomenon occurs due to the
unique property of hyperfine splitted ground state levels. One can
show that for a two-level system with allowed dipole transitions the
pulse will not propagate due to a large absorbtion. The most obvious
way to go out from this situation is the second (coupling) laser
usage that leads to the electromagnetically induced transparency
(EIT, a detailed description see in ref.~\cite{ref.13}).

Now, let us say a few words about the quantity
$\varepsilon_{\text{k}}$, which was neglected in deriving the
equation~(\ref{eq.9}). If we assume $k=(\Delta\varepsilon_{21}/c)$,
one can find that, \textit{e.g.}, for cesium,
${\varepsilon_{\text{k}}\approx3,5\times10^{-30}}$~eV. So, even at
the point $\delta\omega=0$ it is small in comparison with the
linewidth ${\gamma=3,8\times10^{-21}}$~eV. Thus, we proved that the
used approximation is correct.

\section{Pulse slowing concerned with Zeeman splitted hyperfine
         structure levels}
\begin{figure}
\includegraphics{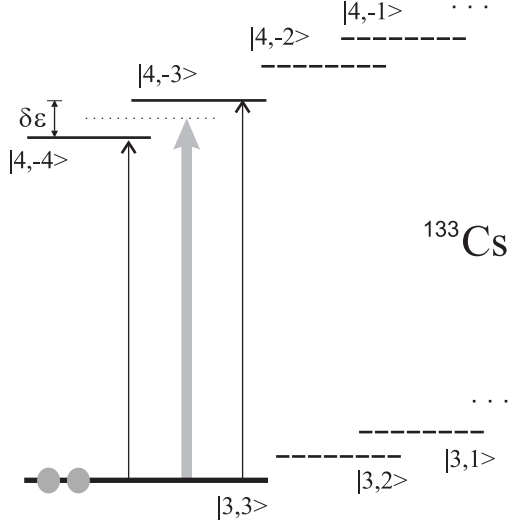} \caption{\small Ground state energy structure of
cesium atom in the external magnetic field. The first sign in
brackets corresponds to the total atomic spin $F$, the second one to
its projection $m_{F}$. For good visibility and compactness not all
sublevels are showed.} \label{fig.1}
\end{figure}

In the previous section the observing possibility for ultra-slow
microwaves caused by hyperfine structure levels was shown. There we
assumed that such levels are degenerate, \textit{i.e.} the external
magnetic field is absent. But the presence of an external magnetic
field (such situation occurs in most of experiments) results in more
complicated picture. Each of hyperfine structure levels splits to
additional components with different total spin projection $m_{F}$
(\textit{e.g.} for cesium atoms see fig.~\ref{fig.1}). As it is well
known, in a weak external magnetic field the energy difference
between such components is proportional to the magnetic field
intensity (the Zeeman effect).

If the components of the upper hyperfine structure level are kept
away from each other (the linewidth of levels is much less than the
energy difference between levels of multiplet~$\delta\varepsilon$,
as shown in fig.~\ref{fig.1}), the signal tuned up exactly to the
transition between the occupied lower state and one of the upper
states (narrow black arrowed lines in fig.~\ref{fig.1}) can be
slowed down, as it is pointed out in the previous section. The only
difference is that the states are no more degenerated
($g_{1}=g_{2}=1$) and the linewidth~$\gamma_{j}$ in eq.~(\ref{eq.7})
corresponds to the transition probability from the chosen component
with the projection $j=(-F',...,F')$ to the occupied lower state.
Taking all other values the same as when deriving the
inequality~(\ref{eq.15}) in the presence of the external magnetic
field one can find:
\begin{equation}\label{eq.16}                              %%%% ---- %%%% Eq. 16
    10^{-1}\text{cm}^{-3}
    \ll|\nu_1 - \nu_2|\ll
    2\times10^{15}\text{cm}^{-3}.
\end{equation}
In the experiment~\cite{ref.14} for the BEC regime cesium atoms with
a peak density of $1,3\times10^{13}~\text{cm}^{-3}$ were kept in a
trap. Thus, according to the inequality~(\ref{eq.16}), one can
conclude that in such system the ultra-slow microwaves phenomenon
could be observed. Moreover, direct calculations show that the group
velocity of the signal propagating in the mentioned system is close
to $9\times10^{-6}$~m/s. It means that the pulse tuned up in such
way covers the distance of 1~millimeter in the condensed cesium
vapour nearly in 2~minutes!

Let us note that to observe such a kind of phenomenon one needs to
use the microwave signal with the linewidth~$\gamma_\text{s}$ much
less than the level linewidth
\begin{equation}\label{eq.17}                              %%%% ---- %%%% Eq. 17
    \gamma_\text{s}\ll\gamma_{j}.
\end{equation}
Therefore, one can meet essential experimental difficulties because,
as it was mentioned earlier, the linewidth concerned with the
transition between hyperfine structure levels is sufficiently small.

However, in the presence of an external magnetic field there are
more regions where slowing down phenomenon can be observed. For
example, the microwave signal, detuned relatively to the transitions
between two neighbour states of the upper hyperfine multiplet and
occupied lower state (wide grey arrow line on fig.~\ref{fig.1}) can
also be slowed down to sufficiently small values. In that case the
group velocity to a greater extent depends on the magnetic field
intensity and the condition~(\ref{eq.17}) is not necessary.

On the basis of the developed approach it is not difficult also to
describe such variant of slowing, taking into account the Zeeman
splitting. If we assume that all atoms are condensed in lower
(occupied) state with the density~$\nu$ and energy~$\varepsilon_0$,
we get (see eq.~(\ref{eq.7})):
\begin{equation}\label{eq.18}                              %%%% ---- %%%% Eq. 18
\begin{split}
    \epsilon^{-1}(\textbf{k},\omega)
    \approx 1&+\dfrac{\nu|\sigma_{12}(\textbf{k})|^{2}}
    {2\pi^2 k^2}
    \\
    &\times\sum\limits_{j=-F'}^{F'}\left[
    \dfrac{1}
    {\omega-\Delta\varepsilon_{j}(H)
    +i\gamma_{j}}\right],
\end{split}
\end{equation}
where $\sigma_{12}(\textbf{k})$ can be found from the
expression~(\ref{eq.14}) and
$\Delta\varepsilon_{j}=(\varepsilon_{|F',m_{F'}>}-\varepsilon_0)$.
Now, using the derived equation, one can study the dependence of the
refractive index on the frequency of microwave signal and its
slowing down conditions.

\begin{figure}
\includegraphics{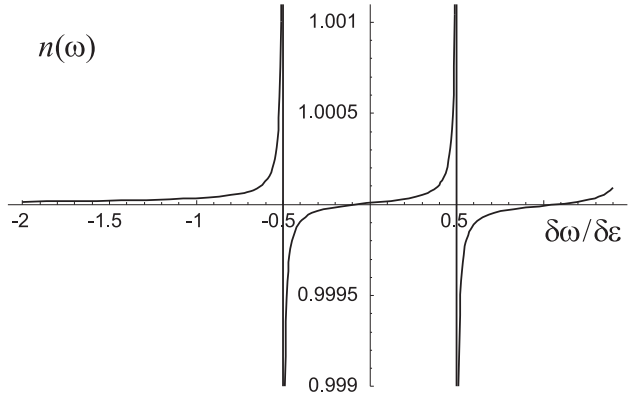} \caption{\small Refractive index dependence for cesium
atoms in BEC state in the region of frequencies close to the
transitions between hyperfine splitted levels. Left and right narrow
peaks correspond to the transitions
$|3,3\rangle\rightarrow|4,-4\rangle$ and
$|3,3\rangle\rightarrow|4,-3\rangle$, respectively. Note that with
frequency increasing (to the right side from the figure edge) the
behavior qualitatively repeats. The used parameter for the Zeeman
splitting energy is $\delta\varepsilon/\gamma_{j}\approx10^3$.}
\label{fig.2}
\end{figure}

Let us demonstrate it also by the cesium atoms example. The
dependence of the refractive index in case of the external magnetic
field presence has a rather complicated form due to a large number
of splitted levels (9 levels in the upper state with $F=4$), thus it
is more convenient to show such dependence in a figure. In
fig.~\ref{fig.2} one can see that in central part ($\delta\omega=0$
that corresponds to the wide grey arrowed line in fig.~\ref{fig.1})
the refractive index has a slope with steepness depending on the
distance $\delta\varepsilon$ between the levels $|4,-4\rangle$ and
$|4,-3\rangle$. It means that the microwave signal, which is detuned
in such way, can also be slowed. The direct calculations show that,
\textit{e.g.} for the ideal gas of cesium atoms in BEC state
($\nu=1,3\times10^{13}\text{cm}^{-3}$,                                   %% CHANGED !!! %%
$\gamma_{j}=3,8\times10^{-21}~\text{eV}$) with the energy of
splitting $\delta\varepsilon=3,8\times10^{-18}~\text{eV}$                %% CHANGED !!! %%
($\delta\varepsilon/\gamma_{j}=10^3$, such as showed on
fig.~\ref{fig.2}), the pulse can be slowed down to 1,07~m/s.

\section{Conclusion}

Thus, by means of the microscopic approach we studied the linear
response of an ideal atomic gas to disturbing effect of an external
electromagnetic field. Our approach was based on a novel formulation
of the second quantization method in the presence of bound states of
particles~\cite{ref.4}. The use of such an approach allowed us to
obtain expressions for the dielectric permittivity of dilute gases
of alkali atoms in BEC state. The existence of resonance frequencies
in expression for permittivity was found.

Our approach raised a possibility to study propagation properties of
the microwave signal, tuned up to the transition between two
hyperfine ground state levels of alkali atoms that were considered
in BEC state. In contrast to dipole-allowed transitions, it was
demonstrated that the pulse could propagate in such system with
rather small energy loss. Due to this fact we introduced the group
velocity concept. The slowing down conditions for the signal that
propagated in BEC (at the limit of zero temperatures) were studied.
Moreover, we revealed the dependence of the group velocity sign on
the level's population difference. Thus, we suggested that in some
cases in a Bose-Einstein condensate could propagate the weak-damping
microwaves with a negative group velocity.

In case of the external static magnetic field (we took into account
the Zeeman splitting) the ultra-slow microwaves phenomenon was also
studied. Considering the example of cesium atoms vapour it was shown
that in some conditions the pulse could be slowed down to extremely
small values.


\begin{thebibliography}{99}

\bibitem{ref.1}
  Hau L., Harris S., Datton Z., and Behvoozi C. Nature \textbf{297}
  (1999), 594.

\bibitem{ref.2}
  Matthews M., Hall D., Jin D., Ensher J., Wieman C.,
  Cornell E., Dalfovo F., Minniti C., and Stringari~S.
  Phys. Rev. Lett. \textbf{81} (1998), 243.

\bibitem{ref.3}
  {Allen L. and Eberly J.}
  \emph{Optical Resonance and Two-Level Atoms}.
  {Dover, New York}
  (1987).

\bibitem{ref.4}
  {Peletminskii S. and Slyusarenko Yu.}
  {J. Math. Phys.} \textbf{46} (2005), {022301}; quant-ph/0605159. %% CHANGED !!! %%

\bibitem{ref.5}
  {Slyusarenko Yu. and Sotnikov A.}
  {Cond. Matt. Phys.} \textbf{9} (2006), {459}; cond-mat/0702637.   %% CHANGED !!! %%

\bibitem{ref.6}
  {Slyusarenko Yu. and Sotnikov A.}
  {Low Temp. Phys.} \textbf{33} (2007), {30}.

\bibitem{ref.7}
  {Akhiezer A. and Peletminskii S.}
  \emph{Methods of Statistical Physics}.
  {Pergamon, Oxford}
  (1981).

\bibitem{ref.8}
  {Akhiezer A., Peletminskii S., and Slyusarenko Yu.}
  {JETP} \textbf{86} (1998), {501}.

\bibitem{ref.9}
  {Clementi E., Raimondi D., and Reinhardt W.}
  {J. Chem. Phys.} \textbf{38} (1963), {2686}.

\bibitem{ref.10}
  {Clairon A., Salomon C., Guellati S., and Phillips W.}
  {Europhys. Lett.} \textbf{12} (1991), {683}.
  %{Tiesinga E., Verhaar B., Stoof H., \and van Bragt D.}
  %{Phys. Rev. A}{45}{1992}{2671}.

\bibitem{ref.11}
  {Gu\'{e}ry-Odelin~D., S\"{o}ding~J., Desbiolles P., and Dalibard J.}
  {Europhys. Lett.} \textbf{44} (1998), {25}.

\bibitem{ref.12}
  {Kadomtsev B.}
  \emph{Collective phenomena in plasmas}.
  {Pergamon, New~York}
  (1978).

\bibitem{ref.Chu82}                                               %% CHANGED !!! %%
  Chu~S. and Wong~S. Phys. Rev. Lett. \textbf{48} (1982), 738.

\bibitem{ref.Macke85}                                             %% CHANGED !!! %%
  Segard B. and Macke~B. Phys. Lett. \textbf{109A} (1985), 213.

\bibitem{ref.13}
  {Harris S.}
  {Physics Today} \textbf{50} (1997), {36}.

\bibitem{ref.14}
  {Weber T., Herbig J., Mark M., N\"{a}gerl H., and Grimm R.}
  {Science} \textbf{299} (2003), {232}.

\end{thebibliography}
\end{document}